\documentclass[twocolumn,floatfix,superscriptaddress,amsmath,showpacs,showkeys,aps,pre,eqsecnum]{revtex4-2}
\usepackage[utf8]{inputenc}
\usepackage{bm}
\usepackage[normalem]{ulem}
\usepackage{times}
\usepackage{verbatim}
\usepackage{hyperref}

\begin{document}
\title{Emergent Gauge Symmetry in Active Brownian Matter}
\author{Nathan Silvano}
\affiliation{Center for Advanced Systems Understanding, Untermarkt 20, 02826 G\"orlitz, Helmholtz-Zentrum Dresden-Rossendorf, Bautzner Landstraße 400, 01328 Dresden, Germany}
\affiliation{Departamento de F{\'\i}sica Te\'orica,20270-004
Universidade do Estado do Rio de Janeiro, Rua S\~ao Francisco Xavier 524, 20550-013,  Rio de Janeiro, RJ, Brazil.}

\author{Daniel G. Barci}
\affiliation{Departamento de F{\'\i}sica Te\'orica,20270-004
Universidade do Estado do Rio de Janeiro, Rua S\~ao Francisco Xavier 524, 20550-013,  Rio de Janeiro, RJ, Brazil.}
\affiliation{Sorbonne Université, Laboratoire de Physique Théorique et Hautes Energies, CNRS UMR 7589, 4 Place Jussieu, 75252 Paris Cedex 05, France}
\date{\today}

\begin{abstract}
We investigate a two-dimensional system of interacting Active Brownian Particles.  Using the Martin-Siggia-Rose-Janssen-de Dominicis formalism,  we built up the generating functional for correlation functions.  We study in detail the hydrodynamic  regime with a constant density stationary state.  Our findings reveal that, within a small density fluctuations regime,  an emergent     $U(1)$ gauge symmetry arises, originated from the conservation of fluid vorticity.  Consequently,  the interaction between the orientational order parameter and density fluctuations can be cast into a  gauge theory, where the concept of ``electric charge density" aligns with the local vorticity of the original fluid.  We study in detail the case of a microscopic local two-body interaction. We show that, upon integrating out the gauge fields, the stationary states of the rotational degrees of freedom satisfy a non-local Frank free energy for a nematic fluid. We give explicit expressions for the splay and bend elastic constants as a function of the P\'eclet number  (${\rm Pe}$) and the diffusion interaction constant ($k_d$).   
\end{abstract}

\maketitle

\section{Introduction}
		Since the introduction of the ``Boids model" by C. W. Reynolds in 1987 \cite{Reynolds_1987} to simulate the swarming of animals, the study of non-equilibrium self-propelled systems has rapidly advanced, largely due to the development of the field of ``active matter".  Active matter refers to a collection of particles that can convert environmental energy into mechanical work, growth, or replication. This field finds numerous applications in biological systems \cite{RevModPhys.91.045004,RevModPhys.88.045006,RevModPhys.85.1143,doi:10.1146/annurev-conmatphys-070909-104101, viswanathan2011physics,Cavagna-2022}. One specific category within active matter is Active Brownian Particles (ABP), which combines Brownian motion and self-propulsion. ABPs are frequently utilized to describe the movement of microorganisms such as bacteria \cite{PhysRevLett.100.218103,PhysRevLett.122.248102} or artificial micro-swimmers such as Janus particles \cite{Walther_Muller_2013}.

Numerous studies have focused on both individual ABPs \cite{RevModPhys.88.045006, Garcia-Millan_2021,PhysRevX.9.021009, PhysRevE.98.062121} and collective properties\cite{PhysRevE.74.061908,PhysRevLett.114.018302,PhysRevLett.114.198301}.
Emergent behaviors, such as clustering \cite{RevModPhys.88.045006,PhysRevLett.117.148002,Zhang_Beer_Florin_Swinney_2010,PhysRevE.65.061106}, and motility-induced phase separation (MIPS) \cite{doi:10.1146/annurev-conmatphys-031214-014710,digregorio2021unified}, continue to garner significant scientific interest \cite{Grobmann_Aranson_Peruani_2020,Redner_2013,PhysRevE.107.014608}. The macroscopic regime in which these emergent behaviors manifest is particularly valuable from a physics perspective,  as it allows one
to look for universal properties  that do not depend on microscopic details. 
As a result, numerous researchers are drawing parallels between active systems and more ``traditional" areas of physics. For instance, parallels with hydrodynamics are readily observed \cite{RevModPhys.85.1143,Julicher_2018}, as well as the use of topology and the emergence of nematic phases \cite{PhysRevX.12.010501}.  Dynamical renormalization group techniques\cite{Nathan-2023}  play a fundamental role in determining universality classes in active matter systems\cite{Cavagna-2021,Cavagna-2021-2,RubioPuzzo-2022} as well as  to study entropy production\cite{Caballero_2018}.   Additionally,  equivalence of active models with quantum dynamics has also been explored in different scenarios\cite{te_Vrugt_2023}.  

Various approaches have been developed to construct coarse-grained models for the study of active Brownian matter.  One of the conventional methods in statistical mechanics is the Dynamic Density Functional theory \cite{doi:10.1080/00018732.2020.1854965}.  Additionally, methods such as the interaction-expansion method enable the systematic microscopic derivation of field theories for systems of interacting active particles \cite{PhysRevResearch.2.033241,Vrugt_2023}.
In this context, the use of coarse-grained models based on the Mori–Zwanzig theory\cite{doi:10.1021/acs.jpcb.1c01120} results in an equation of motion resembling a generalized Langevin equation,  thereby preserving the underlying fine-grained dynamics.  Moreover, the conventional derivation of the Kawasaki and Dean equation \cite{KAWASAKI-1994, DavidSDean_1996}  has been applied to describe active Brownian matter \cite{PhysRevE.103.032607,PhysRevResearch.3.013100,PhysRevLett.100.218103}.
Furthermore,  intriguing approximations, such as those involving infinite-dimensional systems, have yielded exact results regarding these complex systems \cite{Lozano-2019}.

The main goal of this paper is to provide a macroscopic formulation of the interaction between density fluctuations and the orientational degrees of freedom of a two-dimensional system of  interacting ABP's.  We show that,  in a small density fluctuation regime,  there is an emergent gauge symmetry that can be traced back to the conservation of the fluid vorticity.  In this way, the system can be described by a dual $U(1)$ gauge theory, where the emergent  ``electric charge"  corresponds with the vorticity of the original fluid.  The main result of the paper is the gauge theory presented in Eq. (\ref{eq:SGauge}), that describes the interaction between density fluctuations and the orientational degrees of freedom of the self-propelled particles for a model with a general microscopic two-body  potential.

From a technical point of view,  we begin with the usual model of a system of interacting ABPs and built up a generating functional for correlation functions using the Martin-Siggia-Rose-Janssen-de Dominicis (MSRJD) formalism\cite{MSR1973,Janssen-1976,deDominicis}.  Path  integral techniques have been used to represent stochastic models for decades.  However,  they have more recently been applied to describe active Brownian matter\cite{Kiryl-2019,  Banerjee-2022}.   After taking the continuum limit and making a small density fluctuation approximation, we end up with an equivalent  $U(1)$ gauge theory.  We study some properties of local microscopic interactions, explicitly deducing a non-equilibrium thermodynamic potential\cite{BarciMiguelZochil2016}  for the  stationary states. We show that the free energy of the orientational degrees of freedom has the structure of a two-dimensional non-local Frank free energy for a nematic liquid crystal\cite{chaikin-1995,Lubensky-2022,Pearce-2021}, where the bend and splay elastic constants can be cast in terms of the P\'eclet number (${\rm Pe}$)\cite{RAPP2017243} and the diffusion interaction parameter ($k_d$)\cite{connolly2012,Kopp2020}.  

The paper is organized as follows:   in section \ref{S:ABP} we present the  case of a single ABP,  showing the functional formalism for this simple case.   In \S\ref{S:ABM} we address the problem of a system of $N$ ABPs, explicitly constructing  the generating functional.   We describe the hydrodynamics regime and the emergent gauge symmetry in section \ref{S:Hydrodynamic}.  Here,  we display the main result of the paper: the  dual gauge theory describing the interaction of density fluctuations and the orientational degrees of freedom (Eq. \ref{eq:SGauge}).  In \S\ref{S:LP} we show a particular case of local interactions, studying some properties of the steady state.   We discuss our results in section \ref{S:Discussions} and provide some details of the calculations in Appendix \ref{App:MSRJD} and \ref{App:rotdiv}.

\section{Warming up: single active Brownian particle}
\label{S:ABP}
In order to establish the main concepts and to present the functional formalism,  in this section 
we consider a {\em single  active Brownian particle} in two dimensions. Its dynamics is given by the following 
system of overdamped Langevin equations:
\begin{align}
\label{eq:Langevin}
\frac{d {\bf r}(t)}{dt}&= v_0 {\bf n}(t)+\boldsymbol{ \xi}^T(t)\; ,  \\
\label{eq:Langevin-varphi}
\frac{d \varphi(t)}{dt}&= \xi^R(t) \; , 
\end{align}
where ${\bf r}(t)$ is the particle position and the unit vector 
\begin{equation}
{\bf n}(t)=\left(\cos\varphi(t), \sin\varphi(t)\right)\; .
\end{equation}
The  Gaussian white noises $\boldsymbol{ \xi}^T(t)$ and $\xi^R(t)$ are defined by
\begin{align}
\langle \xi^T_\alpha\rangle &=\langle \xi^R\rangle=0 \; ,  \\
\langle \xi^T_\alpha(t)\xi^T_\beta(t')\rangle&=2D_T \delta_{\alpha\beta}\delta(t-t')\; ,  \\
\langle \xi^R(t)\xi^R(t')\rangle&= 2 D_R\delta(t-t') \; .
\end{align}
Greek indexes indicate Cartesian coordinates in the plane $\alpha,\beta=1,2$. Moreover, we are using bold characters for vector quantities.
$D_T$ and $D_R$ are the translation  and rotational diffusion constants respectively. 
For $v_0=0$, the system is a simple Brownian motion, and the diffusion constant can be  identified with the environment's temperature,  $D_T=k_B T$, where $k_B$ is the Boltzmann constant.

Some two point correlation functions can be explicitly  computed.  For instance\cite{ebbens-2010,  Jain-2017}, 
\begin{equation}
\left\langle {\bf v}(t)\cdot {\bf v}(t')\right\rangle=4 D_T \delta(t-t')+ v_0^2 e^{-D_R |t-t'|} \; .
\label{eq:Fluctuations-v}
\end{equation}
where ${\bf v}=d{\bf r}/dt$.  Thus, velocities are correlated for  times smaller than $\tau=1/D_R$.  For a longer time scale the velocities are uncorrelated. 

\subsection{Path Integral  Formalism}

In order to compute higher order  n-point correlation functions  it is convenient to use a path integral representation of the stochastic process. The idea is to built a generating functional 
$Z[\boldsymbol{\eta}^T,  \boldsymbol{\eta}^R]$ depending on vector sources $\boldsymbol{\eta}^T$ and  $\boldsymbol{\eta}^R$  related with translation and rotational degrees of freedom, in such a way that, 
\begin{align}
\left\langle r_{i_1}(t_1)\ldots r_{i_n}(t_n)n_{j_1}(\tau_1)\ldots n_{j_n}(\tau_m) \right\rangle=\nonumber \\ \left.\frac{\delta Z[\boldsymbol{\eta}^T,  \boldsymbol{\eta}^R]}{\delta \eta^T_{i_n}(t_n)\ldots\delta \eta^T_{i_1}(t_1)\delta \eta^R_{j_n}(\tau_m)\ldots\delta \eta^R_{j_1}(\tau_1)}\right|_{\boldsymbol{\eta }^{R,T}=0}\; .
\end{align}

The generating functional is defined as 
\begin{equation}
Z[\boldsymbol{\eta}^T,  \boldsymbol{\eta}^R]=\left\langle
\exp\left\{\int dt \left( \bar{\bf r}\cdot \boldsymbol{\eta}^T+\bar{\bf n}\cdot \boldsymbol{\eta}^R\right)\right\}
\right\rangle\; , 
\end{equation}
where $\{\bar{\bf r},  \bar{\bf n} \}$ is a solution of Eqs. (\ref{eq:Langevin}) and (\ref{eq:Langevin-varphi}) for a given noise configuration.  The brackets $\langle\ldots\rangle$ indicate noise average. 
The difficulty of this expression resides in the fact that an explicit solution of the Langevin equation enters the definition.  In order to avoid this problem,  we introduce a functional integration over two  vectors $\{{\bf r}(t), {\bf n}(t)\}$, in such a way that, 
\begin{align}
Z[\boldsymbol{\eta}^T,  \boldsymbol{\eta}^R]&=\int {\cal D}{\bf r}(t){\cal D}{\bf n}(t) \nonumber \\
&\times\left\langle \delta\left({\bf O}_R\right)\delta\left({\bf O}_T\right) 
\det\left[
\begin{array}{cc}
\frac{\delta {\bf O}_T}{\delta {\bf r}} &  \frac{\delta {\bf O}_T}{\delta{\bf n}} \\
& \\
\frac{\delta {\bf O}_R}{\delta {\bf r}} &  \frac{\delta {\bf O}_R}{\delta {\bf n}}
\end{array}\right] 
\right\rangle_{{\bf \xi_T}, \xi_R} \nonumber \\
&\times  \exp\{ \int dt \left(\boldsymbol{\eta}^T \cdot {\bf r}+\boldsymbol{\eta}^R \cdot \boldsymbol{n}\right)\}  \; .
\label{eq:Z}
\end{align}
In this expression, the sources $\boldsymbol{\eta}^R$ and $\boldsymbol{\eta}^T$ couple with the integration variables $\{{\bf r}(t), {\bf n}(t)\}$ and not with an explicit solution of the Langevin equations.   The $\delta$-functions constrain the integration over this set of solutions.  Explicit expressions of ${\bf O}_{T}$ and ${\bf O}_{R}$ are given by, 
\begin{align}
O^\alpha_{T}&=\frac{d r^\alpha(t)}{dt}- v_0 n^\alpha(t)-\xi^{T,\alpha}(t)\; ,  \\
 O^\alpha_{R}&=\frac{d n^\alpha}{dt}+\epsilon^{\alpha\beta}n^\beta \xi^R \; .
\end{align} 
The determinant in Eq.  (\ref{eq:Z}) is a Jacobian of the functional variable transformation 
$\{{\bf r},{\bf n}\}\to \{\boldsymbol{\xi}^T, \xi^R\}$.   The operators inside the determinant are explicitly given by   
\begin{align}
\frac{\delta O^\alpha_{T}}{\delta r^\beta}&=\delta^{\alpha\beta}\frac{d\delta(t-t')}{dt} \; , \\
 \frac{\delta O^\alpha_{R}}{\delta n^\beta}&=\left\{\delta^{\alpha\beta}\frac{d~}{dt}
 + \epsilon^{\alpha\beta}\xi^R\right\}\delta(t-t')\; ,  \\
 \frac{\delta O^\alpha_{T}}{\delta n^\beta}&=-v_0\delta^{\alpha\beta} \delta(t-t')\; , \\
 \frac{\delta O^\alpha_{R}}{\delta r^\beta}&=0\; .
\end{align}
The presence of a noise term in the determinant is an indication of the multiplicative  character of the stochastic processes\cite{Arenas2012-2}.  However, in this case this term is ``transversal" leading to important simplifications\cite{Arenas2018}. 
 
The final step is to compute the noise averages.  This computation can be done exactly (please see Appendix \ref{App:MSRJD}).  The result is 
\begin{align}
Z[\boldsymbol{\eta}^T,  \boldsymbol{\eta}^R]=&\int {\cal D}{\bf r}(t){\cal D}{\bf n}(t)  \delta\left({\bf n}\cdot{\bf n}-1\right) \label{eq:Z0d}     \\
&\times  \exp\left\{-S[{\bf r}, {\bf n}]+ \int dt \left(\boldsymbol{\eta}^T \cdot {\bf r}+\boldsymbol{\eta}^R \cdot \boldsymbol{n}\right)\right\},
\nonumber
\end{align}
where the action $S[{\bf r}, {\bf n}]$ is given by
\begin{align}
S&=\int dt
\left\{ \frac{1}{4 D_T} \left|\dot{\bf r}\right|^2-\frac{v_0}{2D_T}\dot{\bf r}\cdot {\bf n} +\frac{1}{4 D_R}  \left|\dot{\bf n}\right|^2 \right.  \nonumber \\
&\left.+ \frac{v_0^2}{4 D_T}\right\} \; .
\label{eq:S0d}
\end{align}
Eqs. (\ref{eq:Z0d}) and (\ref{eq:S0d})  are the main result of this section. 
The integration measure in Eq. (\ref{eq:Z0d}) constrains the integration over ${\bf n}$, to a set of unit vectors. The action of Eq. (\ref{eq:S0d}) has a very simple form.   The intrinsic  velocity of the self-propelled particle  
$v_0$ couples the direction ${\bf n}(t)$ with the velocity $\dot{\bf r}(t)$,  in such away that the action is minimized when both quantities are parallel to each other.  The last term of Eq. (\ref{eq:S0d}) is a constant and,  of course,  it does not modify fluctuations.  However,  we decided to keep it since it has a clear physical meaning. It is the entropy production rate,  as we illustrate it below. 
It is important to emphasize that Eqs. (\ref{eq:Z0d}) and (\ref{eq:S0d}) are an exact representation of the system of Langevin equations (\ref{eq:Langevin}) and (\ref{eq:Langevin-varphi}).  In fact, 
it can be checked that the two point correlation function Eq. (\ref{eq:Fluctuations-v}) is exactly reproduced using this formalism.  

The functional formalism is very useful,  not only to compute higher order correlation functions,  but also to study symmetries, conservation laws and  fluctuation theorems. 
A simple example is to look for the entropy production rate.
The heat dissipated into the environment $Q$ , and thus, the increase of entropy in the
medium $\Delta S_m$ associated with a specific trajectory, is given by\cite{Arenas2012-2}
\begin{equation}
\Delta S_m= {\cal T} S[{\rm r}(t),{\rm n}(t)]-S[{\rm r}(t),{\rm n}(t)]\; , 
\end{equation}
where ${\cal T}$ is the time reversal operator.   Using Eq. (\ref{eq:S0d}) we obtain, 
\begin{equation}
\Delta S_m= \frac{v_0}{D_T}\int dt \; \dot{\bf r}\cdot {\bf n} \; .
\end{equation}
We find the total increase of entropy in the medium by averaging over all trajectories 
\begin{equation}
\langle\Delta S_m\rangle=\frac{v_0}{D_T}\int dt \;\langle \dot{\bf r}\cdot {\bf n}\rangle
\; .
\end{equation}
On the other hand,  by using Eqs. (\ref{eq:Langevin}) and (\ref{eq:Langevin-varphi}),  it is simple to deduce   that $\langle \dot{\bf r}\cdot {\bf n}\rangle=v_0$.   In this way,  we  find for the entropy production rate
\begin{equation}
\frac{d\langle\Delta S_m\rangle}{dt}=  \frac{v^2_0}{D_T} \; , 
\end{equation}
that is exactly the last constant term in Eq. (\ref{eq:S0d}).
Thus, this simple model of an active Brownian particle is producing entropy at a constant rate proportional to the square of the  intrinsic velocity of the self-propelled particle.

\section{Multiparticle system}
\label{S:ABM}
In order to study  collective behavior of active particles, we extend the previous model to a system on $N$ interacting  ABPs.
The system of Langevin equations now reads,  
\begin{align}
\frac{d {\bf r}_i(t)}{dt}&= v_0 {\bf n}_i(t)+\boldsymbol{\xi}^T_i(t)-\sum_{j\neq i} {\nabla}U(|{\bf r}_i-{\bf r}_j|))
\; , 
\label{eq:LangevinT} \\
\frac{d n^\alpha_i}{dt}&= -\epsilon^{\alpha\beta}n_i^\beta  \xi^R_i(t)
\; , 
\label{eq:LangevinR}
\end{align}
with ${\bf n}_i\cdot{\bf n}_i=1$.   $U(|{\bf r}_i-{\bf r}_j|))$ is an arbitrary pair potential between particles. 
The noise satisfies 
\begin{align}
\langle \xi^T_{\alpha,i}\rangle &=\langle \xi_i^R\rangle=0 \; ,  \\
\langle \xi^T_{i,\alpha}(t)\xi^T_{j,\beta}(t')\rangle&=2D_T \delta_{ij}\delta_{\alpha\beta}\delta(t-t')\; ,  \\
\langle \xi_i^R(t)\xi_j^R(t')\rangle&= 2 D_R \delta_{ij}\delta(t-t')\; , 
\end{align}
In these equations,  Latin indexes $i,j=1,\ldots, N$ label the particles, while the Greek indexes $\alpha,\beta=1,2$ are  Cartesian components in the plane.   
It is important to stress that  Eqs. (\ref{eq:LangevinT}) and (\ref{eq:LangevinR}) should be interpreted in the Stratonovich stochastic prescription. This is so  because,  only in this prescription the transverse equation 
(\ref{eq:LangevinR}) implies a constant modulus of the vectors ${\bf n}_i(t)$ (Please see Refs. \cite{Aron2014,Arenas2018} and Appendix \ref{App:MSRJD}).

This system of Langevin equations is quite difficult to treat analytically. 
In this paper we present a functional formalism in order to improve our understanding of the mathematical structure of the system.

\subsection{Functional formalism}
The functional formalism for the multiple particle system\cite{Miguel2015} follows the same line of reasoning of the previous single particle case.  In appendix \ref{App:MSRJD},  we  built up  the generating functional of correlation functions of Eqs. (\ref{eq:LangevinT})-(\ref{eq:LangevinR}).  The generating functional can be cast in the form,   
\begin{align}
Z=&\int \left(\prod_i{\cal D}{\bf r}_i(t)\right) \left(\prod_j{\cal D}{\bf n}_j(t)  \delta\left({\bf n}_j\cdot{\bf n}_j-1\right)\right) \\
&\times  \exp\left\{-S[{\bf r}_i, {\bf n}_i]+\sum_i \int dt \left(\boldsymbol{\eta}_i^T \cdot {\bf r}_i+\boldsymbol{\eta}_i^R \cdot \boldsymbol{n}_i\right)\right\}\; ,  \nonumber
\end{align}
where, as before,  $i,j=1,\ldots,N$ labels each particle.  
The first line of this equation is the integration measure where the local constrain $|{\bf n}_j(t)|=1$ is imposed by the delta-function.  In the second line, 
the action can be split into 
\begin{equation}
S=S_T+S_R+S_I\; .
\label{eq:S}
\end{equation}
The translation degrees of freedom are described by
\begin{align}
&S_T= \int  dt \sum_i   \label{eq:ST}    \\ 
&\times \left\{ \frac{1}{4D_T} \left|\frac{d {\bf r}_i(t)}{dt}\right|^2 +\frac{1}{4D_T} \left|\boldsymbol{\nabla}_{r_i}\tilde U\right|^2 
-\frac{1}{2} \nabla_{r_i}^2 \tilde U \right\} \; , 
\nonumber
\end{align}
where 
\begin{equation}
\tilde U({\bf r}_i)=\sum_{j\neq i} U(|{\bf r}_i-{\bf r}_j|) \; .
\end{equation}
$S_T$, given by Eq. (\ref{eq:ST}), describes the dynamic of the passive overdamped Brownian system ($v_0=0$).

On the other hand, the action for  the rotational degrees of freedom is given  by
\begin{align}
&S_R= \frac{1}{4D_R} \int  dt \sum_i \left|\frac{d {\bf n}_i(t)}{dt}\right|^2 \; .
\label{eq:SR}
\end{align}
The active character of the system is codified in the interaction terms 
\begin{align}
&S_I=- \frac{v_0}{2D_T}\int  dt \sum_i \left\{ {\bf n}_i\cdot\frac{d{\bf r}_i}{dt}+
{\bf n}_i\cdot\boldsymbol{\nabla}_{r_i}\tilde U \right\} \; .
\label{eq:SI}
\end{align}
The first term is completely equivalent to the single particle case, favoring configurations in which the actual velocity of each particle is aligned with ${\bf n}_i$.  The second term is proper of a multiparticle system and favors the alignment of ${\bf n}_i$ with the total force felt by the particle $i$, due to the interactions of all other particles. The interaction action $S_I$ is a subtle  balance between both quite different effects. 

Similarly to the single particle case,   this formalism is an exact representation of the system of Langevin equations. In the next sections, we explore some properties of this action in specific approximations.   

\section{Hydrodynamic regime: emergent  U(1) gauge symmetry}
\label{S:Hydrodynamic}
We are interest in a huge number of active particles at finite density, {\em i. e.}, the number of particles $N$ per unit area $A$ is finite, even in the limit $N\to\infty, A\to\infty$. Under this assumption, we  consider a continuum limit that 
we implement it by simply promoting the particle label  $i=1,\ldots,N$ to a continuum two-dimensional vector
${\bf  y}\equiv (y_1,y_2)$.  Moreover, sums over particles turn out to transform into integrals, 
\begin{align}
i &\to  {\bf y}  \; ,  \\
\sum_i&\to \rho_0\int d^2y \; ,  
\end{align}
where $\rho_0$ is a constant density. 
In this way
\begin{align}
{\bf r}_i(t)& \to {\bf r}({\bf y}, t)\; ,  \\
{\bf n}_i(t)& \to {\bf n}({\bf r}(y), t) \; .
\end{align}
The last equation explores the fact that each variable ${\bf n}_i$,  representing the orientation of the velocity $v_0$ of the $i^{\rm th}$ particle,  is glued to the position ${\bf r}_i$
of the particle $i$. 

The passive particle contribution to the action is given by  Eq. (\ref{eq:ST}).  We can re-write it,  in the continuum limit, as  
\begin{equation}
S_T=S_K+S_U\; ,  
\end{equation}
where the kinetic part is given by 
\begin{equation} 
 S_K=
  \frac{\rho_0}{4D_T} \int  dt d^2y\left|\partial_t {\bf r}({\bf y}, t)\right|^2
\label{eq:SK}
\end{equation}
and the  interaction potential contribution is
\begin{align}
S_U&=\frac{\rho_0^3}{4D_T}  \int dt d^2y d^2y' d^2y''  \nonumber \\
& \times \boldsymbol{\nabla} U({\bf r}(y)-{\bf r}(y'))\cdot\boldsymbol{\nabla} U({\bf r}(y)-{\bf r}(y''))
\nonumber \\
& -\frac{\rho_0^2}{2}
\int dt d^2y d^2y' \nabla^2 U({\bf r}(y)-{\bf r}(y')) \; .
\label{eq:SU}
\end{align}
In Eqs.  (\ref{eq:SK}) and (\ref{eq:SU}),  ${\bf r}({\bf y}, t)$ is a vector field of $\{{\bf y},t\}$ and $\partial_t$ stands for the partial derivative with respect to time. 

A key  observation is that $S_T[{\bf r}({\bf y}, t)]$ is invariant under  area preserving diffeomorphisms\cite{Susskind-1991}.  To be concrete,  consider the variable transformation
\begin{equation}
y'_i=y_i+\epsilon_{ij}\partial_j \Lambda({\bf  y}) \; 
\label{eq:APD}
\end{equation}
with  unit Jacobian, 
\begin{equation}
J=\det\left(\partial_j  y'_i\right)=1  \; .
\label{eq:Jacobian=1}
\end{equation}
From  now on,   Latin indexes $i,j=1,2$, labels coordinates in the plane $\{y_1,y_2\}$.  $\Lambda({\bf  y})$ is an arbitrary function and the symbol  $\partial_j\equiv  \partial~/\partial y_j$.   The transformation of Eq. (\ref{eq:APD})  preserves areas in the ${\bf y}$ plane. 

$S_T$,  given by Eqs.  (\ref{eq:SK}) and (\ref{eq:SU}) ,  is invariant under the transformation given by Eq. (\ref{eq:APD}). This fact follows from the observation that each component of the field ${\bf r}({\bf y})$ behaves as a scalar,   {\em i.e. },  
\begin{equation}
 r'_\alpha({\bf y}')=r_\alpha({\bf y}) 
 \label{eq:scalar}
\end{equation}   
and the Jacobian is one, according to Eq. (\ref{eq:Jacobian=1}) (We use Greek index $\alpha=1,2$  to label components of the  position vector).   Equation (\ref{eq:scalar}) implies that the position vector transforms as
\begin{align}
\delta r_\alpha({\bf y})&= r'_\alpha({\bf y})-r_\alpha({\bf y})\nonumber \\
&=-\epsilon_{ij}\partial_i r_\alpha({\bf y})\partial_j \Lambda({\bf y}) +O\left(\Lambda^2\right)\; .
\label{eq:APDx}
\end{align} 

Since the transformation of  Eq. (\ref{eq:APD}) is a symmetry of the system,   there is an  associated conserved quantity  given by No\"ether theorem, 
\begin{equation}
\int d^2y \frac{\delta {\cal L}}{\delta \dot r_\alpha} \delta r_\alpha \; , 
\end{equation}
where ${\cal L}$ is the  Lagrangian density associated with the action $S_T$.  Using Eqs.  (\ref{eq:SK}),  (\ref{eq:APDx}) and integrating by parts we find, 
\begin{equation}
\frac{d~}{dt} \left[  \partial_j \left(\epsilon_{ij}\frac{d r_\alpha}{dt}\partial_i r_\alpha  \right)\right]=0\; .
\end{equation}
Therefore,  due to the invariance under infinitesimal area preserving diffeomorphisms,  the quantity 
\begin{equation}
 \omega(y)\equiv \partial_j \left(\epsilon_{ij}\frac{d r_\alpha}{dt}\partial_i r_\alpha  \right)
 \label{eq:vorticity}
\end{equation}
is a constant.   

In order to shed light on the physical interpretation of $\omega$,  let us integrate this quantity in a bounded region $\Omega$ and apply Gauss law,
\begin{equation}
\int_\Omega d^2y\;  \omega(y)= \oint_{\partial\Omega} \frac{dr_\alpha}{dt} dr_\alpha    \; .
\end{equation}
The right side of this equation is the circulation of the fluid velocity on the closed curve $\partial\Omega$. 
The conservation of the circulation of the velocity is known in fluid mechanics as the {\em Kelvin circulation theorem}\cite{Rieutord-book-2015} and is applied to barotropic fluids subject solely to forces deriving from a potential.   Here, we are showing that the same result can be applied to passive overdamped Brownian particles submitted to interactions given by a two-body potential.  The reason behind this conservation is the invariance of the system under infinitesimal area preserving diffeomorphisms.   In principle,  we do not expect to have this conservation in the case of active particles  since   self-propulsion  is generated by non-conservative forces.  However,  in the next section we will show that, in the special case of weak density fluctuations around a uniform background,  the vorticity is a conserved quantity even in the case of ABPs.

\subsection{Emergent U(1) gauge symmetry}
The  local density of particles, $\rho({\bf r})$,  can be defined by 
\begin{equation}
N= \int  d^2r \rho({\bf r})  \; , 
\label{eq:Nx}
\end{equation}
where $N$ is the number of particles.  By considering the field  ${\bf r}({\bf y})$ as a mapping  from $\{{\bf y}\}\to  \{{\bf r}\}$, we can change the integration variables to obtain
\begin{equation}
N\equiv \int  d^2y\;   \bar\rho({\bf y})= \int  d^2y \; \det[\partial_i r_\alpha] \;  \rho({\bf r}({\bf y})) \; , 
\label{eq:Ny}
\end{equation}
where  $\det[\partial_i r_\alpha]$ is the Jacobian of the transformation ${\bf r}\equiv {\bf r}(\bf y)$.
On the other hand, we have the freedom to choose the coordinates ${\bf y}$ in such a way that the density in 
${\bf y}$-space is a constant, $\bar\rho({\bf y})=\rho_0$.  Then, the real density as a function of ${\bf y}$ can be written as
\begin{equation}
\rho({\bf y})= \rho_0\det\left(\frac{\partial y_i}{\partial r_\alpha}\right) \; , 
\label{eq:density}
\end{equation}
where the determinant is the Jacobian of the inverse transformation ${\bf y}\equiv {\bf y}({\bf r}) $.
In particular, a configuration of constant density $\rho({\bf y})=\rho_0$ is characterized by a unit Jacobian, 
{\em i. e.},    $\partial y_i/\partial r_\alpha=\delta_{i\alpha}$.  Then,  a state of constant density is  described by the configuration 
\begin{equation}
{\bf r}({\bf y})={\bf y}\; .
\label{eq:ConstantConfiguration}
\end{equation}
In order to study the dynamics of small density fluctuations around a constant density $\rho_0$, we can parametrize fluctuations using a vector field ${\bf A}({\bf y}, t)$ as\cite{Susskind-1991} 
\begin{equation}
r_\alpha({\bf y},t)=y_\alpha+\frac{1}{\rho_0}\epsilon_{\alpha\beta} A_\beta({\bf y}, t) \; .
\label{eq:fluctuations}
\end{equation}
With this parametrization, an area preserving diffeomorphism,  such as Eq. (\ref{eq:APD}) or Eq. (\ref{eq:APDx})  is now represented as 
\begin{equation}
A_i({\bf y})\to A_i({\bf y})+\rho_0 \partial_i \Lambda(\bf y) 
\end{equation}
which is a usual  $U(1)$ gauge transformation.
The particle density Eq. (\ref{eq:density}) can be rewritten in this approximation as
\begin{equation}
\rho({\bf y},t)=\rho_0+ \boldsymbol{\nabla}\times {\bf A}({\bf y},t)
\end{equation}
(notice that in two-dimensions $\boldsymbol{\nabla}\times {\bf A}({\bf y},t)$ is a pseudo-scalar quantity).
It is clear that the  density is gauge invariant as it should be, since area preserving diffeomorphisms cannot change the density.   Defining an emergent ``magnetic field''  $B=\boldsymbol{\nabla}\times {\bf A}$, we have
that $B$ represents density fluctuations around a uniform background $\rho_0$,  since
\begin{equation}
\delta\rho({\bf y},t)\equiv\rho({\bf y},t)-\rho_0= B({\bf y},t) \; .
\end{equation}
It is necessary to bare in mind that the $U(1)$ emergent gauge symmetry is appearing in the  small fluctuation regime, where  
$\delta\rho/\rho_0= B/ \rho_0<<1$. 

In this parametrization,  the kinetic term of Eq. (\ref{eq:SK}) takes the form, 
\begin{equation} 
 S_K=
  \frac{1}{4D_T\rho_0} \int  dt d^2y\left|\partial_t {\bf A}({\bf y}, t)\right|^2 \; .
\label{eq:SKA}
\end{equation}
Moreover,  using Eqs. (\ref{eq:vorticity}) and  (\ref{eq:fluctuations}),   we can write the condition of zero vorticity as
\begin{equation}
\omega({\bf y},t)=\frac{1}{\rho_0} \boldsymbol{\nabla}\cdot \partial_t{\bf A}=0 \; .
\label{eq:omegaA}
\end{equation}
We can identify Eqs. (\ref{eq:SKA}) and (\ref{eq:omegaA}),  as the usual electric field action,  complemented with the Gauss law  in the temporal gauge $A_0=0$.     We can incorporate the field $A_0$ as a Lagrange multiplier, in order to   get the Gauss law (zero vorticity) as an equation of motion.  In this way, we can write $S_K$ in an explicit gauge invariant form, 
\begin{equation} 
 S_K=
  \frac{1}{4D_T\rho_0} \int  dt d^2y\left|{\bf E}({\bf y}, t)\right|^2  \; , 
\label{eq:SKE}
\end{equation}
where we have defined  the  emergent  ``electric field"  ${\bf E}=-\boldsymbol{\nabla} A_0-\partial_t {\bf A}$.   By  functional deriving $S_K$ with respect to $A_0$ we obtain the Gauss law $\boldsymbol{\nabla}\cdot {\bf E}=0$, which means that the fluid has zero circulation.  The constant $1/2D_T\rho_0$ is playing the role of the vacuum permittivity $\epsilon_0$. 

Terms containing two-body potential interactions are given by Eq.  (\ref{eq:SU}). 
By expanding ${\bf r}({\bf y})$ using Eq. (\ref{eq:fluctuations}) by keeping only the leading order terms in ${\bf A}({\bf y})$, we can write $S_U$  in terms of density fluctuations, that are already gauge invariant.  We have found for the complete  action $S_T$ the expression 
\begin{align}
S_T=
  \frac{1}{4D_T\rho_0} &\int  dt d^2y\left|{\bf E}({\bf y}, t)\right|^2 
   \label{eq:STGauge}  \\
  +&\int dt d^2 y d^2y' B({\bf y},t) V({\bf y}-{\bf y}') B({\bf y}',t) \; , 
 \nonumber
\end{align}
in which
\begin{align}
V({\bf y}-{\bf y}')&=-\frac{1}{2}\nabla^2_y U({\bf y}-{\bf y}') + 
\label{eq:V} \\
&  +\frac{\rho_0}{4D_T}\int d^2z \boldsymbol{\nabla}_z  U({\bf y}-{\bf z})\cdot \boldsymbol{\nabla}_z  U({\bf z}-{\bf y}')\; .
\nonumber 
\end{align}

Therefore, the action for the translation degrees of freedom in the small density fluctuation approximation is completely equivalent to an ``electromagnetic theory" in which the vacuum permittivity  $\epsilon_0=1/2D_T\rho_0$ and $V({\bf y}-{\bf y}')$ is playing the role of a non-local inverse  permeability $\mu_0^{-1}$.
Let us emphasize that this {\em emergent} $U(1)$ symmetry is not an exact symmetry of the hole system.  It is a  manifestation of area preserving diffeomorphisms in the limit of small fluctuations around a constant density.   This is the actual meaning of term {\em emergent symmetry}.

The active character of the system is codified in the coupling given by Eq. (\ref{eq:SI}).
In the continuum limit  this action takes the form
\begin{align}
S_I=&- \frac{v_0\rho_0}{2D_T}\int  dt d^2y  \; {\bf n}({\bf r}({\bf y}),t)\cdot\partial_t{\bf r}({\bf y},t)
\label{eq:SIcont}\\
&- \frac{v_0\rho^2_0}{2D_T} \int  dt d^2y d^2y' \; 
{\bf n}({\bf r}({\bf y}))\cdot\boldsymbol{\nabla}_r U({\bf r}({\bf y})-{\bf r}({\bf y'}))  \; .
\nonumber 
\end{align}
Expanding ${\bf r}({\bf y})$ using Eq. (\ref{eq:fluctuations}), we find to leading order in ${\bf A}$, 
\begin{align}
S_I=& \frac{v_0}{2D_T}\int  dt d^2y  \; {\bf n}({\bf y},t)\times {\bf E}({\bf y},t)
\label{eq:SIgauge}\\
&- \frac{v_0\rho_0}{2D_T} \int  dt d^2y d^2y' \; 
\left(\boldsymbol{\nabla}\cdot{\bf n}({\bf y})\right)U({\bf y}-{\bf y'})  B({\bf y}')  \; , 
\nonumber 
\end{align}
which is explicitly gauge invariant.   The first line of Eq. (\ref{eq:SIgauge}) is the transverse coupling between the emergent electric field and the vector ${\bf n}$. This term is minimized when the electric field is perpendicular to ${\bf n}$.  The second line,  couples the magnetic field with the divergence of ${\bf n}$.  It is worth to mention that while the coupling of the electric field  is a kinematic effect,  the coupling of the magnetic field is dynamical since it depends on the microscopic two-body potential, as shown in the last term of Eq. (\ref{eq:SIgauge}).

Collecting all terms together and writing the interactions in terms of the gauge fields $\{A_0,{\bf A} \}$, we have the following gauge invariant effective action
\begin{widetext}
\begin{align}
S&=\frac{1}{4D_T\rho_0} \int  dt d^2y\left|{\bf E}({\bf y}, t)\right|^2   +\int dt d^2 y d^2y' B({\bf y},t) V({\bf y}-{\bf y}') B({\bf y}',t)\nonumber  \\
&-\frac{v_0}{2D_T}\int  dt d^2y  \left\{A_0({\bf y}, t) \omega({\bf y}, t) + {\bf A}({\bf y}, t)\cdot {\bf J} ({\bf y}, t) \right\}
+  \frac{\rho_0}{4D_R} \int  dt d^2y \left|\partial_t{\bf n}({\bf y},t)\right|^2\; .
 \label{eq:SGauge}
\end{align}
\end{widetext}
The first line of this equation is the action of the  ``free''  emergent electromagnetic field with the usual terms proportional to  ${\bf E}^2$ and $B^2$. The energy contribution of the magnetic field is non-local,  given by $V({\bf y}-{\bf y}')$ that is related with the microscopic two-body potential through  Eq.  (\ref{eq:V}). Notice that this part of the action describes the dynamic of passive overdamped Brownian particles in the weak density fluctuation regime. 
The last term of Eq. (\ref{eq:SGauge}) is the kinetic energy of the orientational field ${\bf n}({\bf y}, t)$, while the first two terms on the second line describe the couplings between the gauge fields and ${\bf n}({\bf y}, t)$.  These terms, proportional to $v_0$, codify  the actual active character of the system. 

 The ``electric charge density" (coupled with $A_0$) is given by  
\begin{equation}
\omega({\bf y}, t)= \boldsymbol{\nabla}\times {\bf n}({\bf y}, t) \; , 
\label{eq:w}
\end{equation}
while the current density (coupled to ${\bf A}$) has two different contributions, 
\begin{equation}
J_{i} ({\bf y}, t)=\epsilon_{ij}\partial_t n_j({\bf y}, t)+J_{i}^{\rm top}({\bf y}, t) \; .
\label{eq:Jw}
\end{equation}
The first term is of pure of kinematic origin while the second term is a topological current  given by
\begin{equation}
J_i^{\rm top}({\bf y}, t)=\rho_0\epsilon_{ij}\partial_j\int d^2y' \; U({\bf y}-{\bf y}') 
\left( \boldsymbol{\nabla}\cdot {\bf n}\right)({\bf y}', t) \; .
\label{eq:Jt}
\end{equation}
$J^{\rm top}$is topological in the sense that $\boldsymbol{\nabla}\cdot{\bf J}^{\rm top}=0$,  independently of the equations of motion. 
Eqs. (\ref{eq:w}) and (\ref{eq:Jw})  satisfy the continuity equation
\begin{equation}
\partial_t \omega-\boldsymbol{\nabla}\cdot {\bf J} =0 \; , 
\end{equation}
as it should be due to gauge invariance.

Eq. (\ref{eq:SGauge}), together with the definitions of Eqs. (\ref{eq:w})-(\ref{eq:Jt}),  is the main contribution of this paper.  It describes the dynamics of small density fluctuations coupled with  orientational degrees of freedom of  active Brownian particles.  
 ``Electric charge" in this dual gauge theory corresponds with vorticity of the original fluid.  Thus, $\boldsymbol{\nabla}\times {\bf n}$ acts a source of vorticity.   On the other hand,  $\boldsymbol{\nabla}\cdot {\bf n}$ induces a topological vortex current given by Eq. (\ref{eq:Jt}).  

In order to go deeper into specific properties of the system,  it is convenient to detail the two-body potential between particles.  In the next section we study the simplest case of a local two-body potential.

\section{Local potential}
\label{S:LP}
Let us analyze in this section the simplest possible two-body  local interaction between active particles.  
Let us consider
\begin{equation}
U({\bf y}-{\bf y}')=U_0\;  \delta^2\left({\bf y}-{\bf y}'\right) \; , 
\end{equation}
where the constant $U_0$ measure the  intensity of the local potential.  $U_0>0$ produces local repulsion between particles while local attraction is modeled  with  $U_0<0$. 

For this model, the inverse permeability turns out to be local $V({\bf y}-{\bf y}')\sim \nabla^2\delta({\bf y}-{\bf y}')$ and the  action of Eq. (\ref{eq:SGauge}) takes the form
\begin{align}
&S=\frac{1}{4D_T\rho_0}
   \int  dt d^2y\left\{\left|{\bf E}\right|^2 +\rho_0^2 U_0^2 \left(1+\frac{2 D_T}{\rho_0 U_0}\right) \left|\boldsymbol{\nabla}B\right|^2 \right.   \nonumber \\
   &\left.  -2 v_0 \rho_0  \left( A_0 \omega + {\bf A}\cdot {\bf J}\right) +  \frac{\rho_0^2 D_T}{D_R} \left|\partial_t{\bf n}\right|^2 \right\}\; .
   \label{eq:SGaugeLocal} 
\end{align}
The first interesting thing to remark is that in the limit of local interactions,   the magnetic term is proportional to $\boldsymbol{\nabla} B$ and not $B$ itself.  This is a clear consequence of the fact that fluctuations with constant density (constant $B$) do not affect the dynamics.  Therefore,   the energy of magnetic domains is proportional to the domain boundaries rather than the domain bulk.  

In the case of repulsive potentials, the coefficient of $\left|\boldsymbol{\nabla}B\right|^2$ is positive.
Thus,  if $v_0=0$,  the system has a tendency to be  homogeneous,   since density fluctuations are penalized. 
However, for attractive potentials $U_0<0$, there is a critical diffusion constant $D_{Tc}=\rho_0 |U_0|/2$ over which the system gets unstable.  In this regime, the greater the number of domain boundaries  the lower the energy.  Then, there is a clear tendency to pattern formation.  
It is important to stress that this  instability for attractive potentials already exist in passive Brownian particles  and it is not related  with clustering  or MIPS,  observed in active particles for repulsive potentials.  In our context, these effects should be a consequence of  the interaction between the Gauge fields $\{A_0,  {\bf A}\}$ and the orientational degrees of freedom ${\bf n}$ (given by the second line of Eq. (\ref{eq:SGaugeLocal})).   
We address this issue in the next subsections.  In \S\ref{S:weaknoise} we  study this interaction in the weak noise limit,  while in  \S\ref{S:Frank} we focus on the orientational degrees of freedom. 

\subsection{Weak noise limit}
\label{S:weaknoise}
In the weak translation noise regime  $D_T \rho_0/D_R<<1$,  the generating functional is dominated by the classical equations of motion. The differential equations for the gauge fields resembles the Maxwell equations in two dimensions. The ``Faraday law''
\begin{equation}
\boldsymbol{\nabla}\times {\bf E}+\partial_t B=0 \; , 
\end{equation}  
as well as the absence of monopoles are satisfied automatically due to gauge invariance. 
On the other hand, the equivalent equations to the Gauss and Ampere laws read
\begin{align}
&\boldsymbol{\nabla}\cdot {\bf E}= v_0\rho_0 \omega  \; , 
\label{eq:Gauss}\\
&\rho_0^2U_0^2\left(1+\frac{2D_T}{\rho_0 U_0}\right)\epsilon_{ji}\partial_j \nabla^2 B+\partial_t E_i=\rho_0 v_0 J_{i} \; , 
\label{eq:Ampere}
\end{align}
where the charge $\omega$ and the current ${\bf J}$ are given by
\begin{align}
\omega&= \boldsymbol{\nabla}\times {\bf n}\; , 
\label{eq:wlocal} \\
J_{i}&=\epsilon_{ij} \partial_t n_j+\rho_0U_0 \epsilon_{ij}\partial_j
\left( \boldsymbol{\nabla}\cdot {\bf n}\right)\; .
\label{eq:Jlocal}
\end{align}
Eqs. (\ref{eq:Gauss}) and (\ref{eq:Ampere}) describe the electromagnetic field configurations,  provided the charge and current distribution are given. Charge and current are in turn determined by  the orientational  field  ${\bf n}({\bf y}, t)$  according with Eqs. (\ref{eq:wlocal}) and (\ref{eq:Jlocal}). To complete the self consistent equations we need to derive an equation for ${\bf n}({\bf y},t)$.
The problem of minimizing $S[{\bf n}]$ with respect to  the orientational  field is that the components are not independent, since they are constrained by the condition of constant modulus, $\left|{\bf n}({\bf y},t)\right|=1$. To do this,  we introduce a Lagrange multiplier $\lambda({\bf y}, t)$ and minimize the extended action 
\begin{equation}
\tilde S[{\bf n}({\bf y},t)]\equiv S +\int dt d^2y \; \lambda({\bf y}, t)\left(\left|{\bf n}({\bf y},t)\right|^2-1\right)\; , 
\end{equation}
 with respect to the now independent variables $\{n_i,\lambda\}$.  The equation of motion for $n$ is obtained by computing
\begin{align}
\frac{\delta \tilde S}{\delta n_i({\bf y},t)}&=0 \; ,  \\
\frac{\delta \tilde S}{\delta \lambda({\bf y},t)}&=0  \; .
\end{align}
We find, 
\begin{align}
\frac{\rho_0^2 D_T}{D_R} \partial^2_t n_i-\lambda n_i&=v_0 \rho_0\left(\epsilon_{ij}E_j+\rho_0 U_0 \partial_i B\right)\; , 
\label{eq:dndt} \\
n_in_i&=1 \; .
\label{eq:n2=1}
\end{align}
Eqs. (\ref{eq:Gauss}), (\ref{eq:Ampere}), (\ref{eq:dndt}) and (\ref{eq:n2=1}) completely determine the dynamics of the  ``electromagnetic field" or conversely, the orientation  field ${\bf n}$ of the self-propelled particles, in a regime where  $D_T\rho_0/D_R<<1$.   

In the static limit,   the  system of equations for the stationary state  takes the interesting form
\begin{align}
&\boldsymbol{\nabla}\times {\bf E}=0\; , 
\label{eq:RotE} \\
&\boldsymbol{\nabla}\cdot {\bf E}= v_0\rho_0\; \left( \boldsymbol{\nabla}\times {\bf n}\right) \; , 
\label{eq:DivE} \\
&\nabla^2 B= -\frac{v_0}{U_0\left(1+\frac{2D_T}{\rho_0 U_0}\right)}\; \left(\boldsymbol{\nabla}\cdot {\bf n}\right)\; , 
\label{eq:Divn} \\
&n_i=\frac{\left(\epsilon_{ij}E_j+\rho_0 U_0 \partial_i B\right)}{\sqrt{|{\bf E}|^2+\rho_0^2 U_0^2 |\nabla B|^2+2\rho_0 U_0 \boldsymbol{\nabla}B\times {\bf E}}}\; .
\label{eq:nEB}
 \end{align}
The first equation is ``Faraday law" while the second equation is the  Gauss law,  in which $\left( \boldsymbol{\nabla}\times {\bf n}\right)$ acts as  a source of electric field and, as we have stated, is a source of  local vorticity.  In Eq. (\ref{eq:Divn}),  we see how   $\left(\boldsymbol{\nabla}\cdot {\bf n}\right)$ is a source of curvature of density modulations. Finally, Eq. (\ref{eq:nEB}) shows that the unit vector ${\bf n}({\bf y})$ has essentially two components: one of them perpendicular to the electric field and the other one in the direction of magnetic field  gradients.  Due to the fact that ${\bf n}$ has constant modulus,  the system of equations is highly non-linear. 

It is streaking to note that, even-though the interaction potential is local,   Eqs.  (\ref{eq:RotE})-(\ref{eq:nEB}) support solutions with modulations of $B$ (density modulations of the fluid).  These solutions could be related with clustering and/or MIPS,  even for repulsive potentials $U_0>0$.  The underling physics behind this effect is the Gauge coupling  between density fluctuations and the orientation of the self propelled particles.  Another important point is that  this system support solutions with local alignments of the vector ${\bf n}({\bf y})$ and vortex configurations,  produced essentially by a combination of the self-propulsion $v_0$ and local microscopic repulsion, $U_0>0$. The effect of velocity ordering in the absence of any microscopic force that explicitly aligns velocities was recently reported in Ref. \onlinecite{Caprini-2020}, where numerical simulations show the emergence of velocity patterns, aligned or vortex-like domains.  
Fluctuations around the saddle-point solutions,  static as well as  dynamical ones,  produce a much complex structure and  deserve specific detailed studies. In the following subsection we advance a further step in understanding the structure of the velocity order. 

\subsection{Orientational degrees of freedom}  
\label{S:Frank}
It is interesting to investigate how density fluctuations induce orientational order of the self-propelled particles.  For this, we can compute  the effective action for the orientational degrees of freedom by integrating out  the gauge fields.  To do that,  we can fix the Coulomb gauge $\boldsymbol{\nabla}\cdot {\bf A}=0$.   In this gauge,   the contribution from $A_0$ and ${\bf A}$ are decoupled, and can be written in the following form
\begin{align}
&S=
   \int  dt d^2y \left\{ \frac{-1}{4D_T\rho_0} A_0 \nabla^2 A_0 +\frac{v_0}{2D_T} A_0 \omega\right\}  \nonumber \\
      &+ \int  dt d^2y    \left\{\frac{-1}{4D_T\rho_0}  A_i \left[\partial_t^2-\rho_0^2 U_0^2\left(1+\frac{2D_T}{\rho_0 U_0}\right)\nabla^4\right]  A_i\right. \nonumber \\
   &  \left.+\frac{v_0}{2D_T}   A_i J_{i} \right\}+  \frac{\rho_0}{4D_R} \int  dt d^2y \left|\partial_t{\bf n}\right|^2 
\; .   \label{eq:SAlocalU}
\end{align}
The integrations over $A_0$ and $A_i$ are Gaussian and can be done without difficulty. 
We find
\begin{widetext}
\begin{align}
S&=\int d^2yd^2y' dt dt' \;  n_i({\bf y}, t) \left\{-\frac{\rho_0}{4D_R}\delta(t-t')\delta^2({\bf y}-{\bf y}')+\frac{v_0^2}{16 D_T^2} G({\bf y}-{\bf y}',t-t' )\right\} \partial_{t'}^2 n_i({\bf y}',t') \nonumber \\
& -\frac{v_0^2 \rho^3_0 U_0^2}{4D_T} \int d^2yd^2y' dt dt'\;
\left[\boldsymbol{\nabla}\cdot {\bf  n}({\bf y}, t)\right] \nabla^2
G({\bf y}-{\bf y}', t-t')  \left[\boldsymbol{\nabla}\cdot {\bf  n}({\bf y}', t')\right]
 \nonumber \\
&+\frac{v_0^2 \rho_0}{4D_T} \int d^2yd^2y' dt \; 
 \left[\boldsymbol{\nabla}\times {\bf  n}({\bf y}, t)\right] G^0({\bf y}-{\bf y}')
\left[\boldsymbol{\nabla}\times {\bf  n}({\bf y}', t)\right)] \; .
\label{eq:Snlocal}
\end{align}
\end{widetext} 
The first line of Eq. (\ref{eq:Snlocal}) describes the dynamics of the vector field ${\bf n}$.  We clearly see that the first term is local in time,  driven by the rotational noise with damping $D_R$.  On the other hand,  density fluctuations induce a non-local and retarded dynamics driven by the Green function $G({\bf y}-{\bf y}',t-t' )$ that satisfies, 
\begin{align}
\left[\partial_t^2-\rho_0^2 U_0^2\left(1+\frac{2D_T}{\rho_0 U_0}\right)\nabla^4\right]
& G({\bf y}-{\bf y}', t-t')   \\
 & =\delta^2({\bf y}-{\bf y}')\delta(t-t') \; , \nonumber
\end{align}
The second line of Eq.  (\ref{eq:Snlocal}) is also dominated by the same Green function that describes the interaction between sources of the orientational field ${\bf n}$.  Finally,  the last line of Eq.  (\ref{eq:Snlocal}) describes the interaction between two-dimensional vortices (or anti-vortices).  This term is local in time, and the potential is the usual Logarithmic interaction between vortices in two dimensions, since  $G^0({\bf y}-{\bf y}')$ satisfies
\begin{equation}
\nabla^2 G^0({\bf y}-{\bf y}')=\delta^2({\bf y}-{\bf y}').
\label{eq:G0}
\end{equation}

Finally,  let us consider the effective  action of  a stationary state configuration.  For these configurations, the effective action growths linearly in time. The effective action per unit time defines a generalized potential that is equivalent to the equilibrium Gibbs free energy\cite{BarciMiguelZochil2016}.  
We obtain
\begin{widetext}
\begin{align}
F_{\rm NLF}=\frac{\rho_0 D_R}{2}\int d^2yd^2y'  \left\{K_s\left[\boldsymbol{\nabla}\cdot {\bf  n}({\bf y})\right] 
G^0({\bf y}-{\bf y}')  \left[\boldsymbol{\nabla}\cdot {\bf  n}({\bf y}')\right]+K_b  \left[\boldsymbol{\nabla}\times {\bf  n}({\bf y})\right] G^0({\bf y}-{\bf y}')
\left[\boldsymbol{\nabla}\times {\bf  n}({\bf y}')\right]\right\}
\label{eq:Frank}
\end{align}
\end{widetext}
that has the form of a non-local  two-dimensional Frank free energy for a nematic fluid\cite{chaikin-1995,Lubensky-2022,Pearce-2021}.
The first and second term correspond to the energy of splay and bend  deformations respectively. It is worth to note that in two dimensions there are no twist deformations\cite{Lubensky-2022}.  In our model, we find for  the splay and bend elastic constants
\begin{align}
K_b&=\frac{ v_0^2}{2D_R D_T} \; , 
\label{eq:Kb} \\
K_s&= \frac{ v_0^2}{2D_R D_T}\left(\frac{1}{1+\frac{2D_T}{\rho_0 U_0}}\right)  \; .
\label{eq:Ks}
\end{align}
From these equations we can identify two different dimensionless parameters.
The bend elastic constant, Eq. (\ref{eq:Kb}) is related with the so called P\'eclet number that it is generally defined by the radio between the advective  and the diffusive transport rate\cite{RAPP2017243},
\begin{equation}
{\rm Pe}= \frac{L v}{D} \; .
\end{equation}
Here, $L$ is a typical length of the system, $v$ is a typical velocity and $D$ is a typical diffusion constant. 
In terms of the parameters of our model, the P\'eclet number can be defined as 
\begin{equation}
{\rm Pe}\equiv \frac{v_0}{\sqrt{2 D_R D_T}} \; .
\label{eq:Pe}
\end{equation}
Just to make contact with the colloid literature, the rotational diffusion constant is usually chosen  as $D_R\sim D_T/\sigma^2$\cite{digregorio2021unified,Gonnella2015}, where $\sigma$ is the typical radius of the colloid. In our local model, a characteristic  length scale is given by the density, so    $\sigma^2\sim 1/\rho_0$, then our definition of  the P\'eclet number is 
${\rm Pe}\sim  v_0 \rho_0^{-1/2}/D_T$, which coincides with most of the literature. We prefer to consider $D_R$ and $D_T$ as independent parameters and keep the expression of Eq. (\ref{eq:Pe}), since it is more natural in our model.   

On the other hand,  from Eq. (\ref{eq:Ks})  we can identify the dimensionless constant
\begin{equation}
k_d\equiv\frac{\rho_0 U_0}{2D_T} \; , 
\label{eq:kd}
\end{equation}
that is the radio of two characteristic energies of the system:  the microscopic two-body interaction and the typical thermal interaction with the bath. $k_d$ is zero for free particles and its sign characterizes repulsive or attractive interactions, provided $k_d$ is positive or negative respectively. We will show that we can identify $k_d$ with the {\em diffusion interaction parameter}\cite{connolly2012,Kopp2020}. 

With these definitions, the elastic constants simply read, 
\begin{align}
K_b&={\rm Pe}^2\; , 
\label{eq:KbPe} \\
K_s&={\rm Pe}^2\left(\frac{k_d}{1+k_d}\right)\; , 
\label{eq:KsPe}
\end{align}
Notice that the ratio $K_b/K_s$ is completely determined by the diffusion interaction parameter $k_d$. 

Two-dimensional nematic phases with long ranged interactions has been studied before\cite{BaRiSt2013,Mendoza-2017}.
However,  in the present case,  very special properties arise due to  the particular Logarithmic interaction in two dimensions.  Interestingly, in our case, the energy of splay and bend deformations are not independent.  To see this, it is not difficult to deduce the following expression (please see Appendix \ref{App:rotdiv})
\begin{align}
 &\left[\boldsymbol{\nabla}\cdot {\bf  n}({\bf y})\right] 
G^0({\bf y}-{\bf y}')  \left[\boldsymbol{\nabla}\cdot {\bf  n}({\bf y}')\right]+  
\label{eq:weakeq}  \\
&\left[\boldsymbol{\nabla}\times {\bf  n}({\bf y})\right] G^0({\bf y}-{\bf y}')
\left[\boldsymbol{\nabla}\times {\bf  n}({\bf y}')\right]=-\delta({\bf y}-{\bf y}') \; .
\nonumber
\end{align}
This is a weak equality,  in the sense that it is only satisfied inside the $y$ and $y'$ integrals.
Thus, we can write the free energy only in terms of splay or, equivalently, in terms of bend deformations.   By replacing Eq. (\ref{eq:weakeq}) into Eq. (\ref{eq:Frank}) we immediately find, 
\begin{align}
F^{\rm bend}_{NLF}&=\frac{\kappa }{2} \int d^2yd^2y' \nonumber \\
&\times 
 \left[\boldsymbol{\nabla}\times {\bf  n}({\bf y})\right] G^0({\bf y}-{\bf y}')
\left[\boldsymbol{\nabla}\times {\bf  n}({\bf y}')\right]\; , 
\end{align}
or, equivalently, 
\begin{align}
F^{\rm splay}_{NLF}&=-\frac{ \kappa}{2}\int d^2yd^2y'  \nonumber \\
&\times \left[\boldsymbol{\nabla}\cdot {\bf  n}({\bf y})\right] 
G^0({\bf y}-{\bf y}')  \left[\boldsymbol{\nabla}\cdot {\bf  n}({\bf y}')\right] \; , 
\end{align}
where we have ignored additive constant terms. 
The stiffness $\kappa=\rho_0 D_R(K_b-K_s)$ is given by
\begin{equation}
\kappa=\rho_0 D_R \frac{{\rm Pe}^2}{\left( 1+k_d\right)} \; .
\label{eq:stiff}
\end{equation}
This expression can be cast in terms of an effective P\'eclet number ${\rm Pe}^2_{\rm eff}$
where 
\begin{equation}
{\rm Pe}^2_{\rm eff}= \frac{v_0^2}{2 D_R D}
\label{eq:Peeff}
\end{equation}
in which the bare translational diffusion $D_T$ has been replaced by the effective diffusion of the interacting system
\begin{equation}
D=D_T \left(1+k_d\right) \; .
\end{equation}
This last equation is exactly the phenomenological  definition of the diffusion interaction constant $k_d$ that, in perturbation theory, is related with the second virial coefficient $B_2$\cite{connolly2012,Kopp2020}.

A curious property of the free energy of Eq. (\ref{eq:Frank}) is that, if both elastic constant are equal $K_s=K_b$, by using  Eq.  (\ref{eq:weakeq}) we conclude that the free energy $F_{\rm NLF}$ is a constant, independently of the orientation of the director ${\bf n}$. Thus,  in this conditions  the system is in a completely orientational disordered phase.  This case corresponds to the extremely strongly coupled regime $\rho_0 U_0\to \infty$ or $k_d\to \infty$ in such way that the stiffness $\kappa\to 0$. This is in contrast with the local Frank free energy,  in which the equality of the elastic constants leads to the celebrated $XY$ model\cite{Pearce-2021}.

\section{Discussions and conclusions}
\label{S:Discussions}

In this paper,  we have studied some aspects of  active Brownian matter.  Specifically, we have considered a system of overdamped Langevin equations (Eqs. (\ref{eq:LangevinT}) and (\ref{eq:LangevinR}) ) describing a set of two-dimensional $N$  active Brownian particles,  interacting thorough a two-body potential $U({\bf r}_i-{\bf r}_j)$.  The active character of each particle is codified in a self propelled velocity $v_0 \hat{\bf n}_i(t)$ with  constant modulus $v_0$ and random orientations of the unit vector $\hat {\bf n}_i$.

By means of a Martin-Siggia-Rose-Jensen-de Dominicis procedure, we have built a functional representation of the system.  This path integral formalism is useful to compute different kind of correlation functions and, perhaps more important,  to study symmetry properties and the corresponding phase transitions.   We focused in the continuum limit,  to describe the hydrodynamic regime of  active Brownian matter. 

Assuming  a uniform  density background,  we have shown the appearance of an emergent $U(1)$ symmetry for weak density fluctuations.    This symmetry is reminiscent of the invariance under area preserving diffeomorphism of the particle system,  when analyzed in a regime of weak density fluctuations. 
Thus,  local density fluctuations can be parametrized by an emergent ``magnetic field",  $\delta \rho=B$,  while the emergent ``electric field'' has a kinetic origin and implements the conservation of the local vorticity in the form of an ``emergent Gauss Law".  In this way,  the effective action can be cast in a similar way than an electromagnetic theory,  albeit with a non-local permeability that depends on the details of the microscopic  two-body potential $U({\bf r}_i-{\bf r}_j)$.  
The gauge field couples minimally with a conserved current ($\omega$, ${\bf J}$). While the charge $\omega$ is related to fluid vorticity by  $\boldsymbol{\nabla}\times{\bf n}$,   the vortex current ${\bf J}$ contains,  beyond a dynamical term $\partial_t {\bf n}$,  a topological component induced by  $\boldsymbol{\nabla}\cdot{\bf n}$.  The conservation of vorticity in the active system can be traced back to the Kelvin circulation  theorem of fluid mechanics that sets the conservation of the circulation of the velocity.  A priori,   we do not expect this kind of conservation for active fluids due to the fact that self propulsion comes form non-conservatives forces.  However,   we have shown that,  in a regime of weak fluctuations around a constant homogeneous density,  the vorticity is conserved due to the appearance of an emergent Gauge symmetry.

The main goal of the paper was to provide a macroscopic description of the interaction between density fluctuations $\delta \rho({\bf y},t)=\rho({\bf y},t)-\rho_0$ and the local orientational field ${\bf n}({\bf y},t)$.
The main result is the gauge theory of Eq. (\ref{eq:SGauge}) that describes this coupling for any reasonable two-body microscopic potential in a regime in which $\delta\rho/\rho_0<<1$.  

As an example,  we have studied in some detail the particular case of a local two-body potential.  In this case, the gauge theory is also local.   Although this is an extremely simple interaction potential, we have shown that the model could describe two streaking properties of active matter, i. e.   clustering and local velocity alignment.  Clustering could be inferred from possible non-homogeneous solutions of the steady state equations of motion in the weak-noise limit.   We have also studied the velocity alignment structure by   
integrating out density fluctuations.  Interestingly,  the effective free energy  for stationary states has the structure of a non-local Frank free energy for a nematic liquid. The bend and splay elastic Frank constants can be cast in terms of two dimensionless parameters: the P\'eclet number (${\rm Pe}$) and the diffusion interaction constant ($k_d$). In fact,  $k_d$,  that codifies the microscopic interactions,  controls the relative weight of bend and splay contributions to the free energy.  Interestingly, the specific Logarithmic interactions makes the global properties of the non-local Frank free energy very different  from its local counterpart.  Remarkably,   it seems that  the simplest two-body  local potential could be a good model to describe some of the main streaking phenomenology observed in active matter,  in experiments as well as numerical simulations. 

We hope that the present  proposal could be a good starting point to help us to improve our understanding of the dynamics as well as the stationary phases of  active matter.  We leave the study of different kinds of  topological defects supported by the action of Eq. (\ref{eq:Snlocal}),  as well as,  the study of its   phase diagram and complex dynamics for a future presentation. 

\acknowledgments
DGB likes to acknowledge  the {\em Laboratoire de Physique Théorique et Hautes Energies (LPTHE)}  at {\em Sorbonne Université}, where part of this work was completed, for kind hospitality.  We would also like to acknowledge Leticia Cugliandolo and Pablo de Castro for useful comments.  The Brazilian agencies, {\em Funda\c c\~ao de Amparo \`a Pesquisa do Rio
de Janeiro} (FAPERJ), {\em Conselho Nacional de Desenvolvimento Cient\'\i
fico e Tecnol\'ogico} (CNPq) and {\em Coordena\c c\~ao  de Aperfei\c coamento de Pessoal de N\'\i vel Superior}  (CAPES) - Finance Code 001,  are acknowledged  for partial financial support.
CAPES-Print program supported NS through a sandwich doctoral fellowship at CASUS, Germany  and DGB through a Senior Visitant Professor fellowship at  Sorbone University, France. 

\appendix
\section{Generating functional for a system of N active Brownian particles}
\label{App:MSRJD}
Consider the following  system of stochastic equations 
\begin{align}
\frac{d {\bf r}_i(t)}{dt}&= v_0 {\bf n}_i(t)-\sum_{j\neq i} \boldsymbol{\nabla}_{r_i}U(|{\bf r}_i-{\bf r}_j|))+\boldsymbol{\xi}^T_i(t) ,
\label{app:EqT} \\
\frac{d n^\alpha_i}{dt}&= -\epsilon^{\alpha\beta}n_i^\beta \xi^R_i(t) \; , 
\label{app:EqR}
\end{align}
where $i=1,\ldots,N$.  Greek index $\alpha, \beta=1,2$ represent the components of two dimensional vectors.  We use bold letters to indicate two-dimensional vector quantities.  

Eq.  (\ref{app:EqT}) is a system of  overdamped Langevin equations for $N$ active particles with positions ${\bf r}_1,\ldots,{\bf r}_N$,  each one with e persistent velocity $v_0 {\bf n}_i$. The particles interact with a force between pairs  given by 
\begin{equation}
{\bf F}_i(\{{\bf r}\})\equiv \boldsymbol{\nabla}_{r_i} \tilde U_i(\{{\bf r}\}) \equiv  \boldsymbol{\nabla}_{r_i}\sum_{j\neq i} U(|{\bf r}_i-{\bf r}_j|)).
\end{equation}
${\bf F}_i(\{{\bf r}\})$ is  the force exerted by the $j=1,\dots,N-1$ particles on the $i^{\rm th}$ particle  by means of the potential $\tilde U_i(\{{\bf r}\})$.  The notation $\{{\bf r}\}$ means the set of  $N$ particles with positions ${\bf r}_1, \ldots,  {\bf r}_N$.  
The vector white noise $\boldsymbol{\xi}^T_i(t)$ is defined by the correlation functions, 
 \begin{align}
\langle \xi^T_{\alpha,i}\rangle &=0  \; , \\
\langle \xi^T_{i,\alpha}(t)\xi^T_{j,\beta}(t')\rangle&=2D_T \delta_{ij}\delta_{\alpha\beta}\delta(t-t') \; ,  
\end{align}
where $D_T$ is the translation diffusion constant that can be identified with the temperature of the environment  $D_T=k_B T$.   On the other hand, the  noise for the rotational degrees of freedom is given by 
\begin{align}
\langle \xi_i^R\rangle&=0 \; ,  \\
\langle \xi_i^R(t)\xi_j^R(t')\rangle&= 2 D_R \delta_{ij}\delta(t-t')  \; , 
\end{align}
where $D_R$ is the rotational diffusion constant. 

The direction of the persistent velocity ${\bf n}_i(t)$ is governed by a stochastic differential equation given by Eq. (\ref{app:EqR}).  
$\epsilon^{\alpha\beta}$ is the complete antisymmetric Levi-Civita tensor in two-dimensions, that guarantee that the equation is automatically transverse.  Multiplying Eq. (\ref{app:EqR}) by $n^\alpha_i$ we immediately obtain
\begin{equation}
{\bf n}_i\cdot\frac{d {\bf n}_i}{dt}=0 \; .
\label{app:transv}
\end{equation}
In the Stratonovich stochastic prescription this transversality implies that 
\begin{equation}
{\bf n}_i\cdot\frac{d {\bf n}_i}{dt}=\frac{1}{2}\frac{d \left({\bf n}_i\cdot {\bf n}_i\right)}{dt}=0
\label{app:nconstant}
\end{equation}
and therefore $|{\bf n}_i(t)|=$constant.  In any other prescription,  Eqs.  (\ref{app:transv}) and (\ref{app:nconstant}) are not equivalent.  Consequently,  the stochastic evolution,  given by Eq.  (\ref{app:EqR}), does not keep the modulus constant in this case.  If for some reason  we insist in modeling the stochastic evolution of a vector with constant modulus,  in any other prescription other than Stratonovich,  we should modify Eq.  (\ref{app:EqR}) properly\cite{Aron2014,Arenas2018}. 

In the following we build up the generating functional for correlation functions of eqs.  (\ref{app:EqT}) and (\ref{app:EqR}).
The generating functional is given by  
\begin{align}
Z[\boldsymbol{\eta}^T,  \boldsymbol{\eta}^R]&=\int \left(\prod_i{\cal D}{\bf r}_i(t)\right) \left(\prod_j{\cal D}{\bf n}_j(t)  \right) \nonumber \\
&\times\left\langle \delta\left({\bf O}_R\right)\delta\left({\bf O}_T\right) 
\det\left[
\begin{array}{cc}
\frac{\delta {\bf O}_T}{\delta {\bf r}} &  \frac{\delta {\bf O}_T}{\delta{\bf n}} \\
& \\
\frac{\delta {\bf O}_R}{\delta {\bf r}} &  \frac{\delta {\bf O}_R}{\delta {\bf n}}
\end{array}\right] 
\right\rangle_{{\bf \xi_T}, \xi_R} \nonumber \\
&\times  \exp\{ \int dt \left(\boldsymbol{\eta}^T \cdot {\bf r}+\boldsymbol{\eta}^R \cdot \boldsymbol{n}\right)\} \; , 
\label{app:Z}
\end{align}
where $\boldsymbol{ \eta}^T$ and $\boldsymbol{\eta}^R$ are sources to compute correlation functions.  In Eq. (\ref{app:Z})  we have introduced the vector functions
\begin{align}
O^\alpha_{T,i}&=\frac{d r^\alpha_i(t)}{dt}- v_0 n^\alpha_i(t)+ \nabla_{r_i}^\alpha \tilde U_i-\xi^{T,\alpha}_{i}(t) \; ,  \\
 O^\alpha_{R,i}&=\frac{d n^\alpha_i}{dt}+\epsilon^{\alpha\beta}n_i^\beta \xi_{i}^R \; .
\end{align} 
The operators in the determinant are given by 
\begin{align}
\frac{\delta O^\alpha_{T,i}}{\delta r_j^\beta}&=\left\{\delta_{ij}\delta^{\alpha\beta}\frac{d~}{dt}+ \nabla_{r_j^\beta}\nabla_{r_i^\alpha} \tilde U_i\right\}\delta(t-t') \; ,  \\
 \frac{\delta O^\alpha_{R,i}}{\delta n_j^\beta}&=\left\{\delta^{\alpha\beta}\frac{d~}{dt}
 + \epsilon^{\alpha\beta}\xi^R_{i}\right\}\delta_{ij}\delta(t-t')  \; , \\
 \frac{\delta O^\alpha_{T,i}}{\delta n_j^\beta}&=-v_0\delta_{ij}\delta^{\alpha\beta} \delta(t-t') \; , \\
 \frac{\delta O^\alpha_{R,i}}{\delta r_j^\beta}&=0 \; .
\end{align}

The main goal of this formalism is to try to exactly integrate over the noise,  in order to have a representation only  in terms of the trajectories ${\bf r}_i$ and ${\bf n}_i$.  To do this we first  exponenciate the  delta functions  by using  a couple of auxiliary vectors ${\bf A}_T$ and ${\bf A}_R$ in such a way that 
\begin{align}
\delta\left({\bf O}_T\right)&=\int  \left(\prod_i{\cal D}{\bf A}_{T,i}(t)\right) e^{i\int dt\sum_i  A^\alpha_{T,i} O^\alpha_{T,i} } \; , \\
\delta\left({\bf O}_R\right)&= \int\left(\prod_i{\cal D}{\bf A}_{R,i}(t)\right) e^{i\int dt\sum_i  A^\alpha_{R,i} O^\alpha_{R,i} } \; .
\end{align} 
We also exponenciate the determinant by using two sets of independent vector Grassmann variables  $\{ \boldsymbol{\bar\psi}_{T,i},\boldsymbol{\psi}_{T,i}\}$ and $\{ \boldsymbol{\bar\psi}_{R,i}\boldsymbol{\psi}_{R,i}\}$ in such a way that
\begin{align}
&\det\left[
\begin{array}{cc}
\frac{\delta {\bf O}_T}{\delta {\bf r}} &  \frac{\delta {\bf O}_T}{\delta{\bf n}} \\
& \\
\frac{\delta {\bf O}_R}{\delta {\bf r}} &  \frac{\delta {\bf O}_R}{\delta {\bf n}}
\end{array}\right] = \\
&\int  \left(\prod_i{\cal D}\boldsymbol{\bar \psi}_{T,i}{\cal D}\boldsymbol{\psi}_{T,i}{\cal D}\boldsymbol{\bar \psi}_{R,i}{\cal D}\boldsymbol{\psi}_{R,i}\right)\exp\left\{ \int dtdt' \sum_{ij}\right. \nonumber \\
& \left.   \bar\psi_{T,i}^\alpha(t) \frac{\delta O^\alpha_{T,i}}{\delta r_j^\beta} \psi_{T,j}^\beta(t') +\bar\psi_{R,i}^\alpha(t) \frac{\delta O^\alpha_{R,i}}{\delta n_j^\beta} \psi_{R,j}^\beta(t')\nonumber   \right\} \; .
\nonumber
\end{align}
With these tricks, the noises enter   the exponential linearly  an can be exactly integrated. 
Collecting all the noise terms  we have, 
\begin{align}
\left\langle I_{\rm noise}\right\rangle&= \left\langle e^{-i\int dt\sum_i A_{T,i}^\alpha \xi_{T,i}^\alpha}  \right\rangle_{\boldsymbol{\xi}_T} \\ 
&\times  \left\langle e^{\int dt\sum_i  \left(i A_{R,i}^\alpha n_i^\beta+\bar\psi_{R,i}^\alpha\psi_{R,i}^\beta\right)\epsilon^{\alpha\beta} \xi_{R,i}} \right\rangle_{\xi_R} \; .
\end{align}
The mean values can be computed exactly since the integrals over the noises are Gaussian. The result is
\begin{align}
&\left\langle I_{\rm noise}\right\rangle= e^{-\int dt\sum_i  \left\{  D_T\left|{\bf A}_{T,i} \right|^2 +D_R\left( {\bf A}_{R,i}\times {\bf n}_i\right)^2 \right\}}   \\
& \times e^{2i D_R\int dt\sum_i  \left\{ \left({\bf A}_{R,i}\times {\bf n}_i \right)\left(\boldsymbol{\bar\psi}_{R,i} \times \boldsymbol{\psi}_{R,i}\right)+ \left(\boldsymbol{\bar\psi}_{R,i}\times \boldsymbol{\psi}_{R,i}
\right)^2 \right\}}  \; .  \nonumber
\end{align}

The next step is the integration over the Grassmann variables.  Collecting all the terms in Grassmann variables we find, 
\begin{widetext}
\begin{align}
&\int  \left(\prod_i{\cal D}\boldsymbol{\bar \psi}_{T,i}{\cal D}\boldsymbol{\psi}_{T,i}{\cal D}\boldsymbol{\bar \psi}_{R,i}{\cal D}\boldsymbol{\psi}_{R,i}\right) \exp\left\{\int dt \sum_i  \left(\boldsymbol{\bar \psi}_{T,i}\cdot \frac{d\boldsymbol{\psi}_{T,i}}{dt}+\boldsymbol{\bar \psi}_{R,i}\cdot \frac{d\boldsymbol{\psi}_{R,i}}{dt}   \right.\right.  \nonumber \\
&\left. +  \left.  2i D_R  \left[ \left({\bf A}_{R,i}\times {\bf n}_i \right)\left(\boldsymbol{\bar\psi}_{R,i} \times \boldsymbol{\psi}_{R,i}\right)+ \left(\boldsymbol{\bar\psi}_{R,i}\times \boldsymbol{\psi}_{R,i}
\right)^2 \right]\right) + \sum_{ij}\left( \bar \psi^\alpha_{T,i} \psi^\beta_{T,j} \nabla_{r_j^\beta}\nabla_{r_i^\alpha} \tilde U_i  \right)\right\}
 \nonumber    \\
 &=\exp\left\{  \int dt \sum_i  \frac{1}{2} \nabla^2_{r_i}\tilde U_i   \right\} \; .
\end{align}
\end{widetext}
The last result was obtained by using the relations
\begin{align}
\left\langle  \bar \psi^\alpha_{T,i}(t)  \psi^\beta_{T,j}(t')  \right\rangle&= \delta_{ij}\delta^{\alpha\beta} G_R(t-t')\; ,  \\
\left\langle    \bar \psi^\alpha_{R,i}(t)  \psi^\beta_{R,j}(t')  \right\rangle &=\delta_{ij}\delta^{\alpha\beta} G_R(t-t')\; ,  \\
\left\langle  \bar \psi^\alpha_{T,i}(t)  \psi^\beta_{R,j}(t')  \right\rangle&=0 \; , 
\end{align}
where $G_R(t-t')$ is the retarded Green's function of the operator $d/dt$ and $G_R(0)=1/2$ corresponding to the Stratonovich stochastic prescription.

After the  noise and Grassmann variables integration we find for the action, 
\begin{align}
S&= \int dt \sum_i D_T \left|{\bf A}_{T,i}\right|^2 \nonumber \\
& +i A_{T,i}^\alpha\left\{\frac{d r^\alpha_i(t)}{dt}- v_0 n^\alpha_i(t)+ \nabla_{r_i^\alpha} \tilde U_i  \right\} \nonumber \\
&+D_R \left({\bf A}_{R,i}\times {\bf n}_i\right)^2+i A^\alpha_{R,i}\frac{d n^\alpha_i}{dt} - \frac{1}{2} \sum_{j\neq i}\nabla^2_{r_i}U \; .
\label{ap:SA} 
\end{align}
Now, we need to integrate over ${\bf A}_T$ and ${\bf A}_R$. The integral over ${\bf A}_T$ is Gaussian and can be done without any difficulty.  The integral over ${\bf A}_R$ is more interesting  since it hides the constraint on  ${\bf n}_i$.  Let show this explicitly. 
For fix ${\bf n}$, we can choose a local frame to write the components of ${\bf A}_R$ in the following way, 
\begin{equation}
A_{R,i}^\alpha\equiv  a_{\parallel,i} n^\alpha_i + a_{\perp,i} \epsilon^{\alpha\beta} n_i^\beta  \; .
\end{equation}
In terms of these coordinates, the second  line of Eq. (\ref{ap:SA}) reads, 
\begin{equation}
D_R  a_\perp^2 + i a_{\parallel,i} {\bf n}_i\cdot  \frac{d {\bf n}_i}{dt} +i a_{\perp,i} \left({\bf n}_i\times  \frac{d {\bf n}_i}{dt}\right) \; .
 \end{equation}
The integral over $a_{\perp,i}$ is Gaussian and can be done easily.  The result is 
\begin{equation}
\frac{1}{4D_R}  \int dt\sum_i  \left({\bf n}_i\times  \frac{d {\bf n}_i}{dt} \right)^2 \; .
\end{equation}
On the other hand, the integral over $a_{\parallel,i}$ is linear, its integration produce the delta-function
\begin{equation}
\delta\left({\bf n}_i \cdot \frac{d {\bf n}_i}{d t} \right) \; .
\end{equation}
This constraint in the Stratonovich stochastic convention is equivalent to a constant modulus,  $\left| {\bf n}_i\right|=$constant.

Putting all terms together
The action can be cast in the following way, 
\begin{equation}
S=S_T+S_R  \; , 
\end{equation}
where 
\begin{align}
S_T&= \frac{1}{4D_T}\int  dt \sum_i \left\{\frac{d {\bf r}_i(t)}{dt}- v_0 {\bf n}_i(t)+ \boldsymbol{\nabla}_{r_i} \tilde U_i  \right\}^2 \nonumber \\
&+\frac{1}{2} \int dt\sum_i \nabla^2_{r_i} \tilde U(\{ {\bf r}\}) \; , 
\label{app:ST} \\
S_R&=\frac{1}{4D_R}  \int dt\sum_i \left( {\bf n}_i \times \frac{d {\bf n}_i}{dt}\right)^2 \; .
\label{app:SR}
\end{align} 
Equation (\ref{app:ST}) is the Onsager–Machlup action corresponding to the stochastic equation Eq. (\ref{app:EqT}). The last term is the Jacobian of  the variable transformation from $\boldsymbol{\xi}^T$ to ${\bf r}$, in the Stratonovich prescription. For this reason is the $1/2$ at the front of this term. 
Moreover,  Eq. (\ref{app:SR}) is the Onsager–Machlup action corresponding to the stochastic equation Eq. (\ref{app:EqR}). The last term is the Jacobian of  the variable transformation from $\xi^R$ to ${\bf n}$, in the Stratonovich prescription. It is important to note, that this action has the implicit constraint $|{\bf n}_i|=1$.  

It is more convenient to rewrite Eqs. (\ref{app:ST}) and (\ref{app:SR}) is a Lagrangian form. To do this we expand the squares of the the first line of both equations. 
Eq. (\ref{app:ST}) can be rewritten as, 
\begin{align}
S_T&= \int  dt \sum_i  \frac{1}{4D_T} \left|\frac{d {\bf r}_i(t)}{dt}\right|^2 +\frac{1}{4D_T} \left|\boldsymbol{\nabla}_{r_i}\tilde U\right|^2 
-\frac{1}{2} \nabla_{r_i}^2 \tilde U \nonumber \\
&-\int  dt \sum_i  \frac{v_0}{2D_T} {\bf n}_i\cdot\left(\frac{d{\bf r}_i}{dt}+\boldsymbol{\nabla}_{r_i}\tilde U\right) \nonumber \\
& + \frac{v_0^2}{4D_T}(t_f-t_i)+\sum_i \frac{\tilde U_i(t_f)-\tilde U_i(t_i)}{2 D_T} \; .
\label{app:STlagrangian}
\end{align}
The first line of this equation is the usual MSRJD action for a system of particles with two-body potential $U$ and an overdamped dynamics characterized by a diffusion constant $D_T$.  The second line, is the coupling with the orientational degrees of freedom coming from the active part of the dynamics,  and is proportional to $v_0/2D_T$.     The last lines,  are constant terms, coming from the integration of total time derivatives.   While these terms are important to compute some equilibrium properties, they do not affect fluctuations given by the correlation functions of ${\rm r}_i$ and ${\rm n}_i$.

In the same way, we find for the rotational degrees of freedom, 
\begin{align}
S_R&=\frac{1}{4D_R} \sum_i\int dt  \left| \frac{d {\bf n}_i}{dt}\right|^2 \; , 
\label{app:SRh}
\end{align}
where we have explicitly used the condition ${\bf n}_i\cdot{\bf n}_i=1$. 

Putting both terms together (Eqs. (\ref{app:STlagrangian}) and (\ref{app:SRh})), we find the action of Eqs. (\ref{eq:S})-(\ref{eq:SI}). 

\section{Relation between bend and splay contributions to the free energy}
\label{App:rotdiv}
In this appendix we  provide an explicit demonstration  of Eq.(\ref{eq:weakeq}) that is at the stem of the relation between the energy of bend and splay deformations.

Let us begin with the expression
\begin{align}
I_{\rm rot}&\equiv\!\!\int d^2y d^2y'\left[\boldsymbol{\nabla}\times {\bf  n}({\bf y})\right] G^0({\bf y}-{\bf y}')
\left[\boldsymbol{\nabla}\times {\bf  n}({\bf y}')\right] \nonumber \\
&=\!\! \int d^2y d^2y'  \epsilon_{ij}\epsilon_{\ell m} \partial_i n_j(y) G^0({\bf y}-{\bf y}') \partial_\ell' n_m(y').
\end{align}
In the second line, we re-wrote the same expression of the first line in component notation.  $\epsilon_{ij}$ is the complete antisymmetric Levi-Civita tensor,  $\partial_i\equiv \partial/\partial y_i$ and $\partial'_j\equiv \partial/\partial y'_j$ and we are assuming summation over repeated index.
Now, we make use of the tensor relation
\begin{equation}
\epsilon_{ij}\epsilon_{\ell m}=\delta_{i\ell}\delta_{jm}-\delta_{im}\delta_{j\ell} \; , 
\end{equation}
obtaining 
\begin{align}
I_{\rm rot}&=\int d^2y d^2y' \left\{ \partial_i n_j(y) G^0({\bf y}-{\bf y}') \partial_i' n_j(y') \right.\nonumber  \\
&\left.-
\partial_i n_j(y) G^0({\bf y}-{\bf y}') \partial_j' n_i(y')\right\} \; .
\end{align}
Integrating by parts in $y$ and $y'$ we find, 
\begin{align}
I_{\rm rot}&=\int d^2y d^2y' \left\{  n_j(y) \partial_i\partial_i'G^0({\bf y}-{\bf y}')  n_j(y') \right.\nonumber  \\
&\left.-
 n_j(y) \partial_i \partial_j' G^0({\bf y}-{\bf y}')  n_i(y')\right\} \; , 
\end{align}
where we have ignored total derivative terms.

In the next step we take advantage of the translational invariant property of $G^0$ and replace, in the first term 
$\partial_i'\to -\partial_i$ and in the second term, $\partial_i\to -\partial_i'$ and  $\partial_j'\to -\partial_j$. We get, 
\begin{align}
I_{\rm rot}&=\int d^2y d^2y' \left\{  -n_j(y) \nabla^2G^0({\bf y}-{\bf y}')  n_j(y') \right.\nonumber  \\
&\left.-
 n_j(y) \partial_i' \partial_j G^0({\bf y}-{\bf y}')  n_i(y')\right\} \; . 
 \label{app:Ir}
\end{align}

One important point is that 
\begin{equation}
\nabla^2 G^0({\bf y}-{\bf y}')=\delta^2({\bf y}-{\bf y}')
\end{equation}
and $n_i({\bf y})n_i({\bf y})=1$.   
Using this expression in the first line of Eq. (\ref{app:Ir}) and integrating by parts in $y$ and $y'$  in the second line, 
\begin{align}
I_{\rm rot}&=\int d^2y d^2y' \left\{  -\delta^2({\bf y}-{\bf y}') \right.
\nonumber \\ &\left. -
 \partial_jn_j(y)   G^0({\bf y}-{\bf y}') \partial_i' n_i(y')\right\}
 \label{app:Irot}
\end{align}
Rearranging terms in Eq. (\ref{app:Irot}) and coming back to vector notation we finally find, 
\begin{align}
&\int d^2y d^2y'\left\{\left[\boldsymbol{\nabla}\times {\bf  n}({\bf y})\right] G^0({\bf y}-{\bf y}')
\left[\boldsymbol{\nabla}\times {\bf  n}({\bf y}')\right] \right.\\
&\left. +\left[\boldsymbol{\nabla}\cdot {\bf  n}({\bf y})\right] G^0({\bf y}-{\bf y}')
\left[\boldsymbol{\nabla}\cdot {\bf  n}({\bf y}')\right]+\delta^2({\bf y}-{\bf y}')\right\}=0\nonumber 
\end{align}
that it is exactly the relation of Eq. (\ref{eq:weakeq}).

Summarizing, as we have said in the main text,  the  expression of Eq. (\ref{eq:weakeq}) only has sense inside the integration over $y$ and $y'$ and essentially  depends on the fact that the interaction is Logarithmic and the modulus of the vector
${\bf n}({\bf y})$ is a constant. 


%

\end{document}